\documentclass[final,5p,times,twocolumn]{elsarticle}

\usepackage{hyperref}
\usepackage{amssymb}
\usepackage{tabularx}
\usepackage{multirow}
\usepackage{bm}
\usepackage{amsmath}
\usepackage{algorithm2e}
\usepackage{placeins}
\usepackage{color}

\journal{HEDP}

\begin{document}

\begin{frontmatter}



\title{Counter-propagating radiative shock experiments on the Orion laser and the formation of radiative precursors}

\author[IC]{T. Clayson}
\ead{thomas.clayson10@imperial.ac.uk}

\author[IC]{F. Suzuki-Vidal}
\author[IC]{S.V. Lebedev}
\author[IC,livermore]{G.F. Swadling}
\author[LERMA]{C. Stehl\'{e}}
\author[IC]{G. C. Burdiak}
\author[AWE]{J. M. Foster}
\author[AWE,FLF]{J. Skidmore}
\author[AWE]{P. Graham}
\author[AWE,livermore]{E. Gumbrell}
\author[AWE,livermore]{S. Patankar}
\author[targets]{C. Spindloe}
\author[LERMA,ELI beams]{U. Chaulagain}
\author[ELI]{M. Kozlov\'{a}}
\author[LPP]{J. Larour}
\author[LERMA,Palaisseau]{R.L. Singh}
\author[lasPalmas]{R. Rodriguez}
\author[lasPalmas]{J. M. Gil}
\author[lasPalmas]{G. Espinosa}
\author[Madrid]{P. Velarde}
\author[AWE,IC]{C. Danson}

\address[IC]{Blackett Laboratory, Imperial College London, SW7 2BW, United Kingdom}
\address[LERMA]{LERMA, Sorbonne Universités, UPMC Univ. Paris 06, Observatoire de Paris, PSL Research University, CNRS, F-75252, Paris, France}
\address[AWE]{AWE Aldermaston, Reading RG7 4PR, United Kingdom}
\address[targets]{Target Fabrication Group, Central Laser Facility, Rutherford Appleton Laboratory, Harwell Science and Innovation Campus, Didcot OX11 0QX, UK}
\address[ELI]{ELI beamlines, Insitute of Physics ASCR, Na Slovance 1999/2, Prague, 182 21, Czech Republic}
\address[Palaisseau]{Ecole Polytechnique, Palaisseau, France}
\address[LPP]{LPP, CNRS, Ecole polytechnique, UPMC Univ Paris 06, Univ. Paris-Sud, Observatoire de Paris,
Universit\'{e} Paris-Saclay, Sorbonne Universit\'{e}s, PSL Research University, 4 place Jussieu, 75252
Paris, France}
\address[lasPalmas]{Universidad de las Palmas de Gran Canaria, Spain}
\address[Madrid]{Universidad Politecnica de Madrid, Spain}

\address[livermore]{Current address: Lawrence Livermore National laboratory, California 94550, USA}
\address[FLF]{Current address: First Light Fusion, United Kingdom}
\address[ELI beams]{Current address: ELI Beamlines, Prague, Czech Republic}

\begin{abstract}
We present results from new experiments to study the dynamics of radiative shocks, reverse shocks and radiative precursors. Laser ablation of a solid piston by the Orion high-power laser at AWE Aldermaston UK was used to drive radiative shocks into a gas cell initially pressurised between $0.1$ and $1.0 \ bar$ with different noble gases. Shocks propagated at {$80 \pm 10 \ km/s$} and experienced strong radiative cooling resulting in post-shock compressions of { $\times 25 \pm 2$}. A combination of X-ray backlighting, optical self-emission streak imaging and interferometry (multi-frame and streak imaging) were used to simultaneously study both the shock front and the radiative precursor. These experiments present a new configuration to produce counter-propagating radiative shocks, allowing for the study of reverse shocks and providing a unique platform for numerical validation. In addition, the radiative shocks were able to expand freely into a large gas volume without being confined by the walls of the gas cell. This allows for 3-D effects of the shocks to be studied which, in principle, could lead to a more direct comparison to astrophysical phenomena. By maintaining a constant mass density between different gas fills the shocks evolved with similar hydrodynamics but the radiative precursor was found to extend significantly further in higher atomic number gases ($\sim$$4$ times further in xenon than neon). Finally, 1-D and 2-D radiative-hydrodynamic simulations are presented showing good agreement with the experimental data.

\end{abstract}


\end{frontmatter}


\section{Introduction}
\label{Introduction}
The effects of radiation on shock dynamics are of interest to many areas of High Energy Density Physics (HEDP) and astrophysics. These radiative shocks are formed in hypersonic flows (Mach number $\gg 1$) where the radiative flux is non-negligible and plays an important role in the structure of the shock \cite{DrakeBook,ZeldovichBook}. Radiative shocks are present in inertial confinement fusion implosions \cite{Pak2013} and numerous astrophysical phenomena, which can be studied by the means of laboratory-astrophysics experiments (see e.g.~\cite{Remington2006}).

At high shock velocities (e.g. $>10 \ km/s$ in neon at $1 \ mg/cc$) radiation flux dominates energy transport in the shock \cite{DrakeBook,ZeldovichBook,Drake2005}. The loss of energy through radiation leads to strong radiative cooling and thus compressions greater than the ideal gas non-radiative limit ($\times 4$ for a monatomic gas). This can result in the formation additional effects, such as the Vishniac thin-shell overstability \cite{Vishniac1983,Laming2002,Grun1991} and thermal cooling instabilities \cite{Meerson1996,Hohenberger2010}. In addition, radiation propagating upstream (into pre-shocked material) can heat and ionize material, resulting in the formation of a “radiative precursor” ahead of the shock \cite{DrakeBook}.

Previous experiments on radiative shocks have been performed with high-power lasers. For example, Sedov-Taylor radiative blast waves can be generated with spherical symmetry by focusing lasers on a pin embedded in a gas \cite{Edens2010}, and cylindrical symmetry by focusing the laser onto cluster gases \cite{Hohenberger2010,Edwards2001,Moore2008}. Cylindrically converging radiative shocks have also been produced by magnetic pressure using pulsed-power machines \cite{Burdiak2015}.

Similarly, a large number of radiative shock experiments have also used high-power lasers to ablate solid materials to act as pistons. These pistons are then able to drive shocks in a gas cell (e.g \cite{Drake2011} and references therein) usually filled with low pressure xenon or low density foams \cite{Foster2005,Keiter2002}. By restricting the transverse width of the gas cell, the shocks act as quasi-one dimensional shocks and can interact with `wall shocks' \cite{Drake2011,Doss2010}. Many of these experiments focused on studying the radiative precursor \cite{Bozier1986,Bouquet2004,Reighard2006,Koenig2006,Stehle2010,Diziere2011,Chaulagain2015} while modifications to this experimental configuration have allowed for the study of more complex phenomena, such as the formation of reverse radiative shocks { \cite{Falize2012} \cite{Krauland2013} \cite{Cross2016}} or collisions with obstacles \cite{Hansen2007,Rosen2009}.

{The experiments detailed in this paper introduce further modifications which expand on the concept of radiative shocks driven in gas cells. A significantly more complex system with two similar counter-propagating, collisional radiative shocks is introduced. The interaction of two identical radiative shocks, as presented in these experiments, is a model for the reflection of both hydrodynamics and radiation off a perfectly reflective surface, providing a unique platform for numerical validation and laboratory astrophysical models. While the collision of two independent shocks is a rare astrophysical event, the formation of reverse shocks, which bare many similarities with these experiments, are common place. These can occur, for instance, when supernovae remnants interact with dense molecular clouds \cite{McKee1974}, within the bow shock of jets launched from young stars \cite{Hartigan2005} and as accretion shocks formed by material falling onto young stars \cite{Orlando2013} or dense objects in cataclysmic variable systems \cite{WarnerBook}.}

These experiments also introduce a new gas cell design with a large transverse width. This allowed shocks to expand freely into a large 3-D volume of gas without being confined by the side walls and effected by `wall shock', found at shock velocities $>60 \ km/s$. This allows for 3-D effects to be studied, which may give rise to perturbations not found in 1-D and 2-D simulations. In principle, this could lead to a more direct comparison to astrophysical phenomena. In addition, the experiments aimed to investigate radiative shock dynamics in a variety of different noble gases between $0.1$ bar and $1.0$ bar. A wide range of diagnostics allowed this experiment to simultaneously study both the post shock and radiative precursor regions of the shock.

Section \ref{experimental set-up} outlines the experiment and the various diagnostics employed. Section \ref{results} presents results from experiments in neon at an initial mass density of $0.49 \pm 0.01 \ mg/cc$ ($0.60 \pm 0.01 \ bar$), and estimates of several shock parameters including shock velocity (\ref{velocity}), post-shock compression (\ref{post shock}) and reverse shock compression (\ref{reverse shock}). The post-shock temperature and ionization are derived from simple models (\ref{precursor}), and comparisons between radiative precursors in different noble gases are presented in \ref{different gases}. Finally, section \ref{sims} presents results from simulations performed prior to (\ref{2D sims}) and after the experiment (\ref{1D sims}) and then compares shock parameters to experimentally determined values (\ref{sims compare}).

\section{Experimental set-up}
\label{experimental set-up}

The experiments produced two similar counter-propagating radiative shocks in a variety of noble gases. This was achieved by focusing high powered lasers onto plastic disks on opposite sides of a gas cell. This resulted in ablation of the plastic disks, driving them forward as pistons into the gas cell and producing a shock.

\subsection{Gas cell targets}
\label{gas cell targets}
\begin{figure*}[t]
	\includegraphics[width=\textwidth]{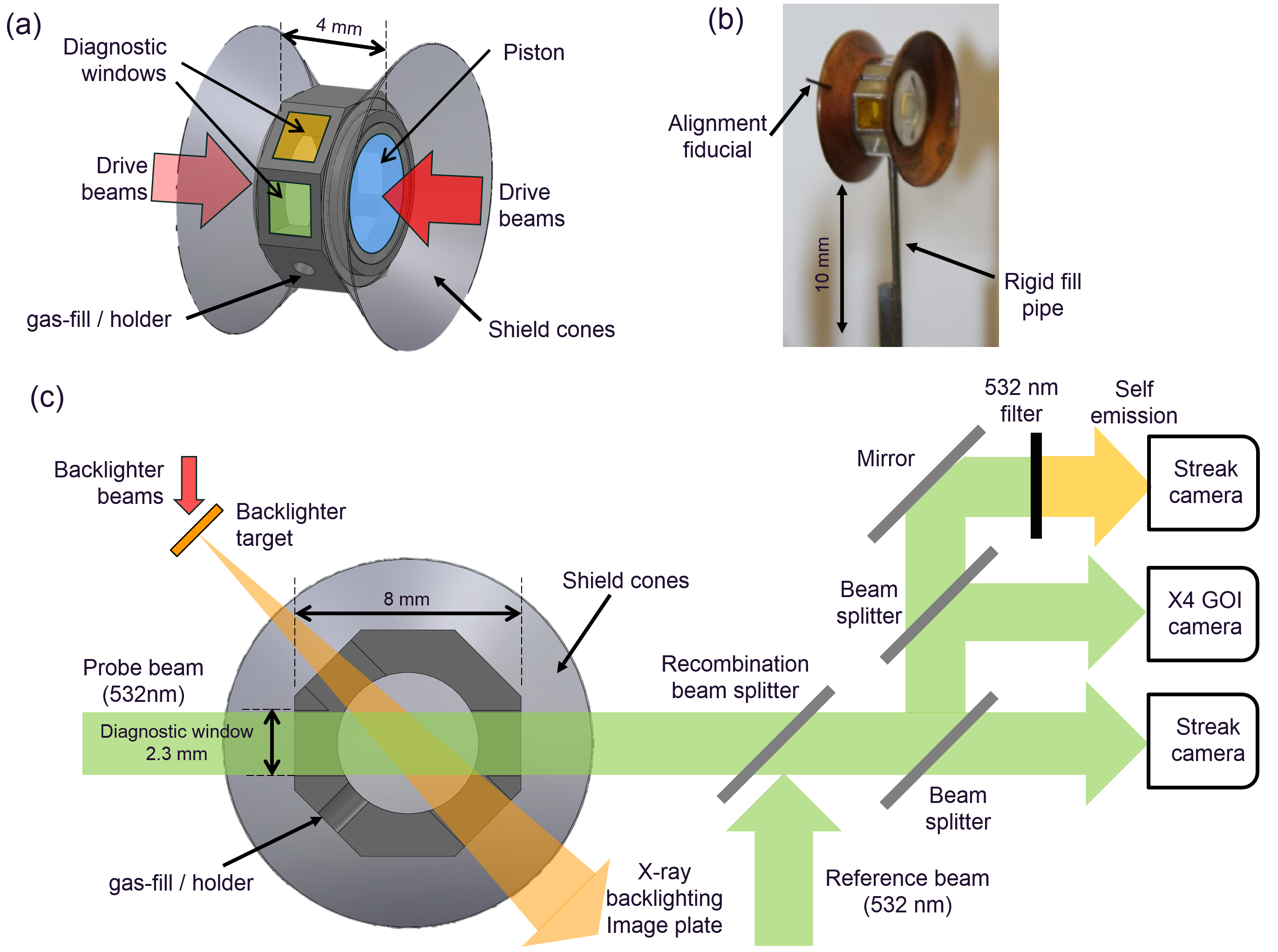}
	\caption{Details of the gas cells used in the experiments (a) 3-D view. (b) Photo. (c) Cross section of the gas cell and schematic of diagnostics fielded on these experiments. This set-up allowed all optical diagnostics to be fielded on the same line of sight.}
	\label{fig:experiment}
\end{figure*}

The gas cells used in the experiments are shown in Fig \ref{fig:experiment}(a-b). To drive radiative shocks, the Orion facility's “long-pulse” beams (1 ns square pulse with a wavelength of $351 \ nm$ \cite{Hopps2015}) were used as drive beams. Four beams delivered a total of $1520 \pm 97 \ J$ to $\sim$$600 \ \mu m$ diameter focal spot on each piston (shown in blue in Fig \ref{fig:experiment}.a), achieving intensities of $\sim$$6 \times10^{14} \ W/cm^{2}$. Pistons were made of polypropylene disks ($5 \ mm$ diameter, $25 \ \mu m$ thick and a density of $\sim$$0.9 \ g/cc$) located on either side of the gas cell, shown in blue in Fig \ref{fig:experiment}.a. Laser ablation of the pistons can result in a significant emission of X-rays and fast electrons. To prevent this emission preheating the gas, a layer of brominated polypropylene (C8H7Br, $3 \ mm$ diameter, $50 \ \mu m$ thick and a density of $\sim$$1.53 \ g/cc$) was attached to the piston on the inside surface of the gas cell. Copper shielding cones surrounded the pistons, shielding the diagnostics from emission from the piston-laser interaction. In addition, a $100 \ \mu m$ wire was attached to these cones, to act as a positioning fiducial for alignment within the Orion vacuum chamber.

The gas cells octagonal bodies were micro-machined from a single piece of aluminium. A $5 \ mm$ diameter hole was drilled through the centre of the octagonal faces and sealed on both ends with the pistons. This wide transverse width, $5 \ mm$, compared to small focal spot, $600 \ \mu m$, allowed the shocks to expand freely into a large volume, avoiding interaction with the gas cell walls \cite{Drake2011,Doss2010}. Four diagnostic windows ($2 \ mm$ by $2.3 \ mm$) were milled onto the rectangular faces and sealed with gas tight filters, shown in yellow and green in Fig \ref{fig:experiment}.a. The windows offered a wide view of the interaction region, with $1 \ mm$ wide regions directly ahead of the pistons obscured by the body of the gas cell.

The gas cells were held in position by a rigid metal gas fill pipe, shown in Fig \ref{fig:experiment}.b. Prior to the experiment, gas cells were filled with noble gases (neon, argon, krypton or xenon) between $0.1$ and $1.0 \ bar$, whilst inside a separate vacuum vessel. The gas cells were then removed from the vessel and exposed to atmosphere while being transferred to the Orion target chamber. The filters and pistons on the gas cells were therefore designed to hold both positive and negative pressures of up to $\sim$$1 \ bar$. While inside the Orion target chamber a pressure transducer was connected to the fill pipe, allowing the gas pressure to be monitored until less than a minute before firing. As a result of pressuring the gas cells, the pistons were found to stretch by $\sim$$100 - 300 \ \mu m$. This was taken into account for alignment of the drive beam lasers to ensure that the focal spot diameter remained $\sim$$600 \ \mu m$ during all experiments.

\subsection{Diagnostic set-up}
\label{diagnostic set-up}
The experiments were diagnosed with point projection X-ray backlighting, laser interferometry and optical self-emission, shown in Fig \ref{fig:experiment}.c.

Point projection X-ray backlighting (XRBL) was used to create a 2-D, time resolved image of the shock through two $25 \ \mu m$ thick polyimide filters attached to the gas cell, acting as windows. To create a bright X-ray source, two backlighter beams ($0.5 \ ns$ square pulse, total energy $\sim$$440 \ J$, wavelength of $351 \ nm$ and synchronised to within $70 \ ps$ \cite{Hopps2015}) were focused to a $400 \ \mu m$ spot on a metal foil, acting as a backlighter target, $21.21 \ mm$ away from the gas cell centre. The resulting X-ray emission is dominated by helium-alpha transitions, resulting in a quasi-monochromatic, narrow band emission spectrum \cite{Matthews1983}. XRBL target material, and thus the X-ray energy, was selected so that variation in ionization of the gas medium would have minimal impact on the transmission, and thus this diagnostic is mostly sensitive to variations in mass density. The XRBL target material was also selected for optimal contrast by comparing synthetic radiographs of simulations (detailed in section \ref{sims}). For experiments with neon Sc XRBL target were used.

The point projection XRBL setup consisted of a $20 \ \mu m$ diameter tantalum pinhole placed over the backlighter target foil. This was coated with a layer of parylene-N to prevent plasma filling the pinhole and absorbing the X-rays. An image, with a magnification of $10.8$, was formed on an image plate (BAS-TR FUJIFILM \cite{Meadowcroft2008} - characterised in \cite{Fiksel2012,Boutoux2016}) $228.6 \ mm$ from the gas cell. The image plate was filtered with two layers of $12.5 \ \mu m$ Ti filters for Sc XRBL targets. To prevent stray light from compromising the image, a light-tight filter of $8 \ \mu m$ thick aluminised polypropylene was placed over the image plate. The spatial resolution was measured to be $27 \pm 5 \ \mu m$ using the X-ray attenuation profile at the sharp edge of the window. In addition, for the shock velocities of $\sim$$80 \ km/s$ (measured in Section \ref{velocity}), the backlighter beams used to generate the X-ray source (with a pulse length of $0.5 \ ns$) result in $\sim$$40 \ \mu m$ of motion blur.

The optical diagnostics, laser interferometry and optical self-emission, were fielded through the same line of sight, as shown in Fig \ref{fig:experiment}.c, through two fused silica filters ($250 \ \mu m$ or $500 \ \mu m$ thick) attached to the gas cell as windows. Laser interferometry was fielded in a Mach-Zehnder configuration with a $532 \ nm$ wavelength probe laser ($50 \ ns$ pulse length with $300 - 400 \ mJ$ and a $\sim$$35 \ mm$ diameter beam). This was imaged with four Gated Optical Intensifiers (GOIs), which recorded 2-D, time resolved interferometry images on the same shot at four different times. The time evolution of a 1-D lineout of interferometry, through the centre of the gas cell and along the axis of shock propagation, was recorded over $100 \ ns$ by an optical streak camera. A 1-D profile of optical self-emission, along the same line, was also recorded over $100 \ ns$ with an additional streak camera, with the $532 \ nm$ probe beam filtered out.

\section{Results}
\label{results}
This section presents results predominantly from experiments in neon at an initial mass density of $0.49 \pm 0.01 \ mg/cc$, i.e. a gas-fill pressure of $0.60 \pm 0.01 \ bar$ at room temperature.

\subsection{Measurements of the shock velocity}
\label{velocity}
\begin{figure}
	\includegraphics[width=0.5\textwidth]{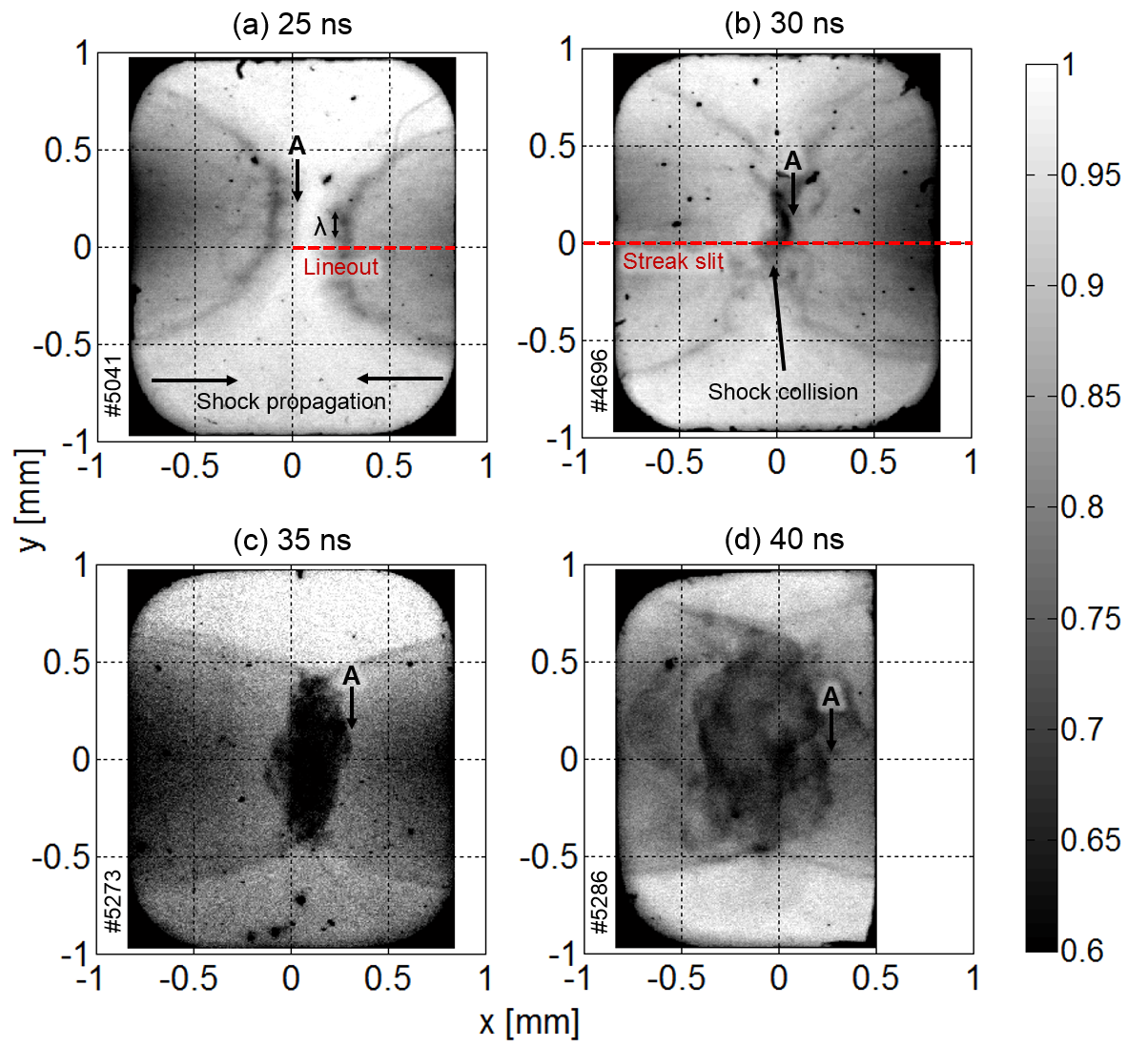}
	\caption{Time sequence constructed from several normalised XRBL images of different shots in neon. The final image (d) was cropped possibly due to diagnostic misalignment. The origin is at the gas cell centre and the pistons are located at $x \sim -2.2 \ mm$ and $x \sim 2.2 \ mm$. The dark dots in the images are a result of debris hitting the image plates.}
	\label{fig:XRBL}
\end{figure}

The shock velocity was measured using several different methods, initially from the position of the shock in XRBL images. Each experiment produced a single XRBL image and a time sequence was constructed from four different experiments, shown in Fig \ref{fig:XRBL}. For these experiments, a Sc XRBL target was used, producing X-rays with an energy of $ \sim$$4.3 \ keV$. Each image shows intensity, normalised to the nominal intensity through the unshocked gas (thus darker regions indicate higher density), through the $2 \ mm$ by $2.3 \ mm$ window (scaled to accounting for point projection effects). The pistons are located at $x \sim -2.2 \ mm$ and $x \sim 2.2 \ mm$, and the two shocks can be seen as semi-circular structures, approaching from the left and right in Fig \ref{fig:XRBL}.a, before colliding near the gas cell centre, at $x = 0$ (shown in Fig \ref{fig:XRBL}.b), $ \sim$$30 \ ns$ after the drive beams.

\begin{figure}
	\includegraphics[width=0.5\textwidth]{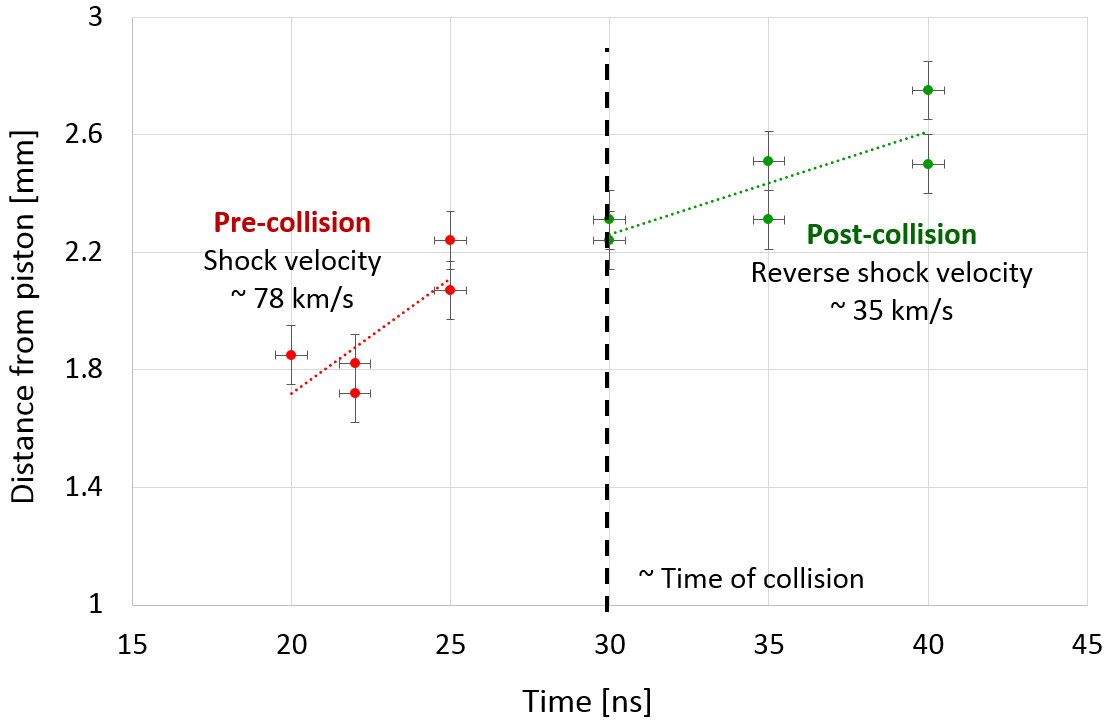}
	\caption{Position of the shock front tip in XRBL images from shots in neon, a selection of which are shown in Fig \ref{fig:XRBL}. The points have been separated into pre-collision and post-collision, and both groups approximated by linear trend lines. The gradient of these trend lines was used to estimate the shock velocity and the reverse shock velocity respectively.}
	\label{fig:XRBLvelocity}
\end{figure}

An initial, rough estimate of the shock velocity was calculated by measuring the position of the shock tip (defined as the edge of the shock, labelled as A in Fig \ref{fig:XRBL}) and dividing by the time the image was taken. This yielded an average velocity of  $\sim$$85 \ km/s$, prior to collision. Characterisation of the gas cells performed prior to the experiments found that the pistons bulged by  $\sim$$200 \ \mu m$ with a $0.6 \ bar$ gas fill, which has been included in these estimates. This velocity is likely an overestimate of the shock velocity at the time the XRBL images were taken because the shocks are expected to decelerate in their early evolution, as they expand into the large 3-D volume, and thus may be travelling slower when they enter the field of view of the gas cell windows. To improve on this estimate, and find the shock velocity just prior to collision, the position of the shock front was plotted against time between $20 \ ns$ and $30 \ ns$, as shown in Fig \ref{fig:XRBLvelocity}. The velocity was determined from the gradient of the best linear fit to these points, yielding a shock velocity of {$78 \pm 17 \ km/s$}.

\begin{figure}
	\includegraphics[width=0.5\textwidth]{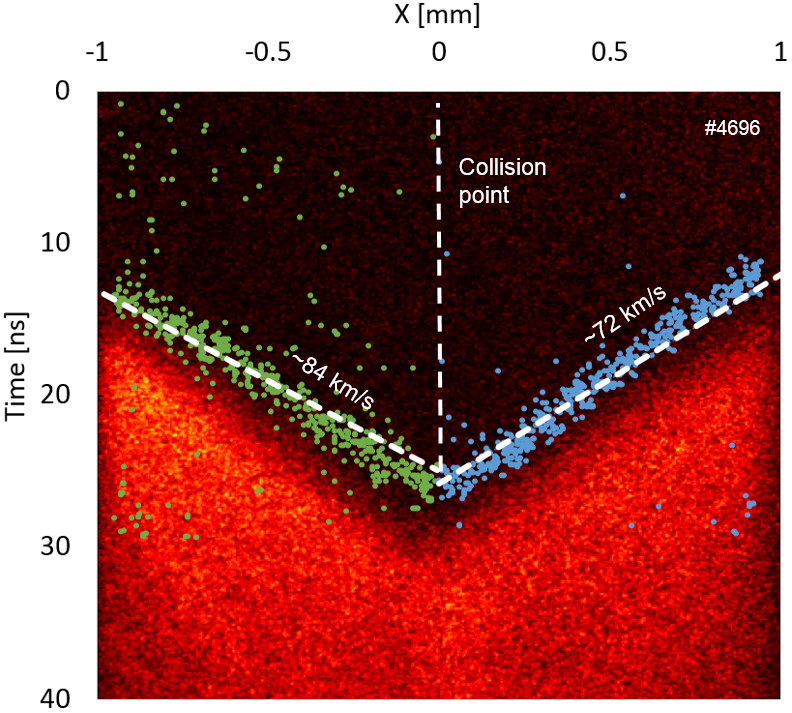}
	\caption{Self-emission streak image along the axis for a shot in neon (XRBL shown in Fig \ref{fig:XRBL}.b). A 1-D slit along the shock axis of propagation was imaged over $100 \ ns$, with spatial position horizontally, time vertically and intensity recorded as pixel brightness.}
	\label{fig:SelfEmission}
\end{figure}

The shock velocity was also measured using optical self-emission streak images. The streak camera recorded a 1-D lineout along the shock axis of propagation, labelled as ``Streak slit'' in Fig \ref{fig:XRBL}.b. Fig \ref{fig:SelfEmission} shows an optical self-emission streak image from a shot in neon (XRBL image shown in Fig \ref{fig:XRBL}.b), with spatial position in the horizontal axis and time in the vertical axis. The shocks can be seen as regions of bright emission that enter the window ($\sim$$1 \ mm$ from the pistons) $\sim$$14 \ ns$ after the drive lasers and propagate towards the collision point in the centre, at $x = 0$, by $\sim$$30 \ ns$. The gradient of the emission edge in the optical self-emission streak image was used to determine the shock velocity. However, the shock front is not well defined, possibly due to emission from the radiative precursor and the broadband wavelength range of the streak camera. To systematically and reliably identify the edge of the shock front, pixels within a range of intensity values were isolated (shown in green and blue in Fig \ref{fig:SelfEmission}). These points were well represented by linear trend lines, the gradient of which was used to determine the shock velocity to be {$80 \pm 10 \ km/s$}, in good agreement with the estimates from the XRBL images.

The consistent shock velocity from both diagnostics allows the Mach number to be estimated, {$M = u/c = 229 \pm 7$}, where $u$ is the shock velocity and $c$ is the sound speed in the cold, unshocked neon,  {$350 \pm 4 \ m/s$}. However, the local Mach number is expected to be significantly lower due to of the existence of a preheated radiative precursor directly ahead of the shock front which should increase the temperature and thus the local sound speed.

Furthermore, XRBL images allow the velocity of the reverse shocks to be estimated. The shocks produced in the experiments are collisional plasmas, with ion mean free path lengths estimated to be $\sim$$1 \ nm$ with formula found within \cite{NRL}. Therefore the post-shock region quickly achieves local thermodynamic equilibrium (LTE) and particles are not able to pass through the interaction region. After the shocks collide, material stagnates in the centre resulting in formation of reverse shocks, shown in Fig \ref{fig:XRBL}.c-d, which propagates through post-shock neon and piston material. The positions of the reverse shock tip were plotted in Fig \ref{fig:XRBLvelocity} and was well approximated with a linear regression line. The gradient of this line yielded a reverse shock velocity of {$35 \pm 12 \ km/s$}.

\subsection{Determining the post-shock compression}
\label{post shock}
The post-shock compression is defined as the ratio of the post-shock mass density, $\rho_{s}$, to the initial unshocked mass density, $\rho_{a}$. For an adiabatic shock in an ideal, monatomic gas with an adiabatic index of $\gamma = 5/3$, the post-shock compression is limited to $\times 4$, (see e.g. \cite{DrakeBook,ZeldovichBook}). However, at high Mach numbers other effects such as radiative losses and ionization can lower the effective adiabatic index, resulting in higher compressions \cite{Michaut2004}. In addition, strong radiative loses can lead to rapid cooling of the post-shock material which, in order to maintain pressure balance, results in further compression behind the shock front.

\begin{figure}
	\includegraphics[width=0.5\textwidth]{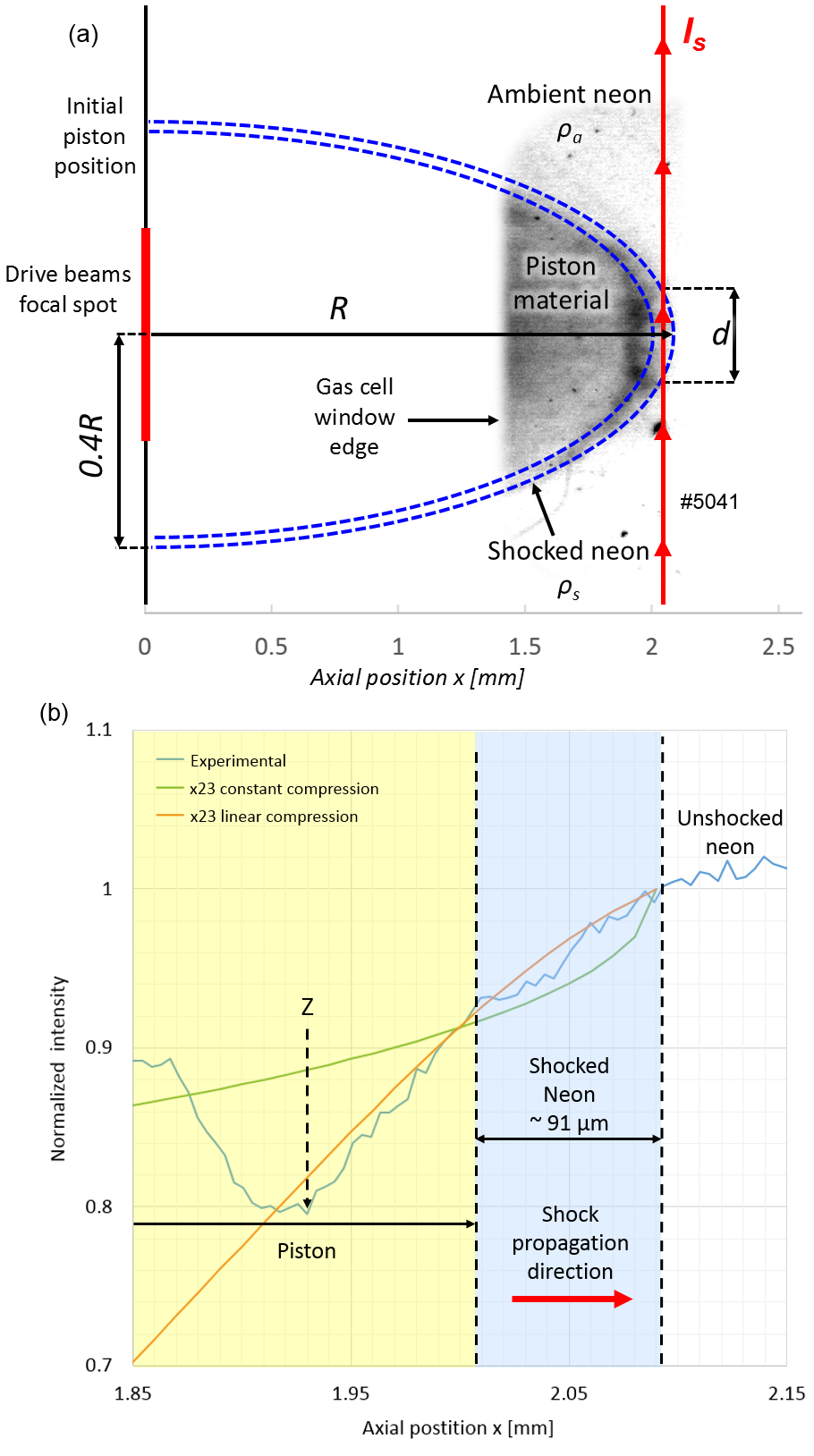}
	\caption{Transmission through the right shock in $0.49 \ mg/cc$ of neon, $25 \ ns$ after the drive laser. (a) The shock was found to be well approximated as a semi-ellipsoid centred on the drive beams focal spot at, $x = 0$, with a semi-minor axis $0.4$ times the semi-major axis, $R$. X-rays, at a position x from the drive beam focal spot, pass predominantly through unshocked ambient neon and a region of post-shock neon, with a length $d$. Due to cylindrical symmetry along the axis, rays coming out of the page experience similar attenuation. (b) Profile of normalized intensity between $x = 1.85 \ mm$ and $x = 2.15 \ mm$ along the shock propagation axis. Transmission was calculated and plotted assuming a constant compression ratio and a linear compression. The shocked neon is highlighted in blue (for an average compression of $\times 23$) and the piston material (C8H7Br) in yellow.}
	\label{fig:PostShockCompress}
\end{figure}

Results of XRBL were used to estimate the compression in the post-shock gas, as this diagnostic is sensitive to changes in mass density. However, the shock front cannot be resolved by the XBRL diagnostic, as it is expected to be of the order of the mean free path ($\sim$$1 \ nm$) \cite{ZeldovichBook} and below the XRBL spatial resolution, $27 \pm 5 \ \mu m$. The XRBL image shown in Fig \ref{fig:PostShockCompress}.a (the right side shock in Fig \ref{fig:XRBL}.a at $25 \ ns$) will be used for the following discussion. The shape of the post-shock region was well approximated as a semi-ellipsoid centred on the drive beams focal spot, at $x = 0$ (obscured by the gas cell body), with a semi-minor axis $0.4$ times the semi-major axis, defined as $R$. This post-shock region is composed of shocked neon followed by piston material (C8H7Br), indicated in blue on Fig \ref{fig:PostShockCompress}.a.

To identify different regions of the shock, a profile of normalised intensity was taken along the axis of the XRBL image. The profile was averaged over $10$ pixels ($\sim$$43 \ \mu m$) and is shown between $x = 1.85 \ mm$ to $x = 2.15 \ mm$ in Fig \ref{fig:PostShockCompress}.b. The unshocked ambient neon is identified on the right, with a normalised intensity of  $\sim$$1$, followed by a region of decreasing intensity from $x \sim 1.93 \ mm$ to $2.09 \ mm$, identified as shocked material. The gradient of intensity within this region is consistent with cylindrical symmetry around the shock axis, with X-rays closer to the piston passing through more shocked material and experiencing additional attenuation. In addition, radiative cooling in the post-shock region could lead to further increases in mass density and thus a steeper gradient. This region consists of shocked neon followed by shocked piston material (C8H7Br). The trough in intensity at $x \sim$$ 1.9 \ mm$ also seen in simulations performed prior to the experiments, and is believed to be composed entirely of piston material. However, the position of the boundary between neon and piston material is initially unclear. The following analysis to determine this boundary will assume negligible mixing between the neon and piston material. 

A first approximation of the average post-shock compression at this time can be estimated by assuming the mass density of post-shock neon is constant. The compression is yielded by dividing the distance the shock has travelled, $\sim$$R = 2.09 \pm 0.01 \ mm$, by the width of the shock region. An upper bound for the shocked neon width was taken to be $0.16 \ mm$, the distance between the unshocked neon with normalised intensity $\sim$$1$ and the trough at $x \sim$$1.93 \ mm$ (labelled as $Z$ in Fig \ref{fig:PostShockCompress}.b). This yielded a lower bound for the post-shock compression of $\sim$$\times 13$. This is higher than the compression limit of an adiabatic shock, $\times 4$, or for a radiation dominated shock, $\times 7$ as shown in \cite{Bouquet2000}. This suggests that ionization and radiative cooling play a significant role in this system.

To more accurately approximate the post-shock compression, the XRBL profile was compared to calculated transmissions. These were calculated using the Beer-Lambert law, which states that monochromatic X-rays, with intensity $I_{0}$, passing through the unshocked neon with a mass density $\rho_{a}$ and length $L = 8 \ mm$, will be attenuated to an intensity $I_{a}$, where $\sigma$ is the mass absorption cross section for neon at $300 \ K$ and $A$ is a constant attenuation due to any filters: $I_{a} = I_{0} A exp(-\sigma \rho_{a} L)$. Fig \ref{fig:PostShockCompress}.a shows how X-rays attenuated by the post-shock neon will have passed through predominantly unshocked neon and a region of post-shock neon with a length $d$. Assuming the post-shock neon has a constant mass density of $\rho_{s}$ the final normalised intensity of the attenuated X-rays will be:

\begin{equation}
	\centering
	\frac{ I_{s} }{ I_{a} } = exp \left[ \sigma d (\rho_{a}-\rho{s}) \right]
\end{equation}

Photon energy of the XRBL was specifically selected so that ionization would have minimal impact on the final transmission, and so the mass absorption cross section, $\sigma = 0.171 \ cm^{2}/mg$ \cite{Henke1993}, is assumed to be constant between the post-shock and unshocked regions. The length of the path through the post-shock region, $d$, can be calculated as a function of the axial position, $x$, assuming the shock can be modelled as a semi-ellipsoid: $d = 0.8 \sqrt{ R^{2}-x^{2} }$. However, plotting this transmission for a constant compression (shown in Fig \ref{fig:PostShockCompress}.b) yielded a curve which does not well represent the profile from the XRBL image, shown in green in Fig \ref{fig:PostShockCompress}.b. This is likely because the mass density is expected to increase through the post-shock region due to radiative cooling. To improve upon this estimate and include this effect, the  post-shock mass density, $\rho_{s}$, was modelled to be linearly increasing behind the shock front, this was approximated as the weighted average between the minimum and average mass densities, $\rho_{min}$ and $\rho_{average}$.

\begin{equation}
	\centering
	\rho_{s} = \frac{R-x}{W} \rho_{average} + \left( 1 -\frac{R-x}{W}  \right) \rho_{min}
\end{equation}

Where $W$ is the post-shock regions thickness, calculated to be the distance the shock has travelled divided by the compression (under the assumption that on the axis the shock acts 1-dimensionally). The minimum mass density was taken to be the shock compression limit for an adiabatic shock, $\times 4$, multiplied by the ambient mass density, $\rho_{a}$ (however, the resulting transmission profile is not strongly dependant on this choice). The average mass density was taken to be the post-shock compression multiplied by the ambient mass density.

The transmission was calculated for several post-shock compressions to find which best approximated the experimental transmission over the width of the post-shock region, shown in Fig \ref{fig:PostShockCompress}.b. A compression of {$\times 23\pm 2$} was found to well approximate the transmission for the shock, shown in orange. A similar analysis found a compression of {$\times 27 \pm 2$} best approximated the transmission for the other shock in this experiment, left side of Fig \ref{fig:XRBL}.a.

However, the shocks are not fully 1-D and thus this estimate is less valid away from the axis. To improve upon the accuracy of this estimate, numerical models are required. Furthermore, laser-target XRBL sources can produce a significant background of hard X-rays ($ > 10 \ keV$) due to fast electrons interacting with the pinhole material \cite{Fein2014,Krauland2012}. The $25 \ \mu m$ titanium filter placed over the image plate was significantly transparent to X-rays above $9 \ keV$ and so these background X-rays could have a significant effect on the previous estimates. To reliably use XRBL images to infer densities, a method of accurately characterising these hard X-rays is required, as discussed in \cite{Fein2014,Krauland2012}.

In summary, the average post-shock compression ($25 \ ns$ after the drive beams) was estimated to be { $\times 25 \pm 2$}, corresponding to an average post-shock mass density of $12 \pm 1 \ mg/cc$. This compression is comparable to radiative shock experiments in 1-D shock tubes with xenon gas fills \cite{DrakeBook,Stehle2012}.

\subsection{Determining the reverse shock compression}
\label{reverse shock}
\begin{figure}
	\includegraphics[width=0.5\textwidth]{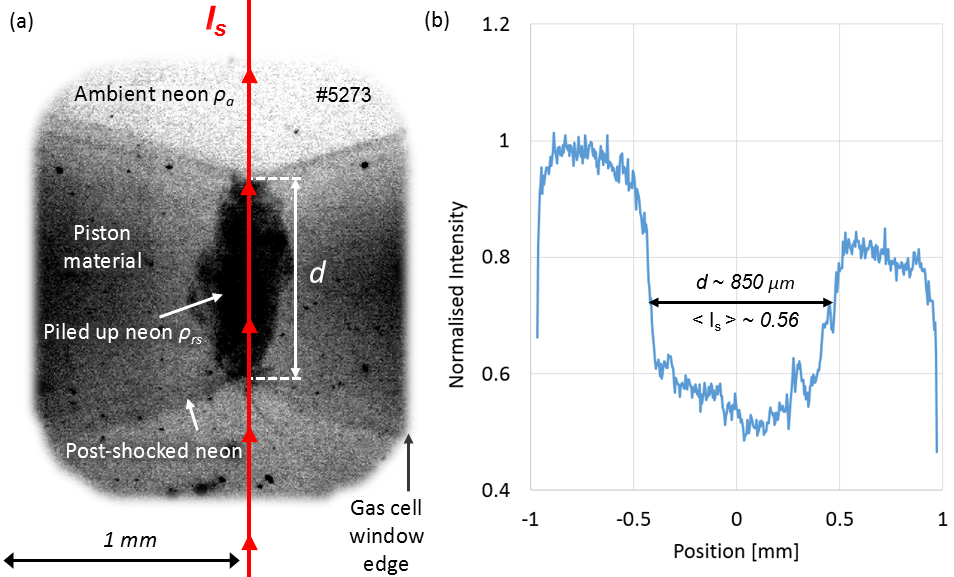}
	\caption{(a) XRBL image $35 \ ns$ after the drive lasers in neon at $0.49 \ mg/cc$, from Fig \ref{fig:XRBL}.c. The two shocks from either side have collided resulting in material stagnating in the centre with a density of $\rho_{rs}$. X-rays passing through this region are attenuation over the length $d$, resulting in a final intensity of $I_{s}$. (b) lineout of normalized intensity along the central axis, showing the width of the stagnated material and average intensity within this region.}
	\label{fig:reverseShock}
\end{figure}

After the shocks collide ($\sim$$30 \ ns$ after the drive beams) a structure forms at the gas cell centre, visible in XRBL images (see Fig \ref{fig:XRBL}.c-d). The shocks in this experiment are collisional and unable to pass through one another, instead material stagnates at the centre of the gas cell and forms two reverse shocks, shown in Fig \ref{fig:reverseShock}.a.

The mass density of material in the post-reverse shock region, $\rho_{rs}$, was measured directly from the intensity in the XRBL images within this region, assuming cylindrical symmetry. Fig \ref{fig:reverseShock}.a shows how X-rays passed through predominantly ambient neon and the post-reverse shock region with a length $d$. In a similar manner to the previous section, the material within the post-reverse shock region is assumed to be of relatively constant mass density and composed predominately of neon, $\rho_{rs}$ with a constant mass absorption cross section between regions, $\sigma = 0.171 \ cm^{2}/mg$, the final intensity of the attenuated X-rays will be $I_{rs}$ given by:

\begin{equation}
	\centering
	I_{rs} = I_{0} A exp \left[ -\sigma \rho_{a} (L-d) \right]  exp \left[- \sigma \rho_{rs} d \right]
\end{equation}

By normalizing this with X-rays passing solely through the unshocked plasma, an expression for the mass density within the post-reverse shock region can be derived:

\begin{equation}
	\centering
	\rho_{rs} = \rho_{a} - \frac{ ln \left( I_{rs}/I_{a} \right) }{ d \sigma }
\end{equation}

Fig \ref{fig:reverseShock}.b shows a lineout of normalised intensity through reverse shock region and the centre of the gas cell (averaged over $10$ pixels, { $\sim$$43 \ \mu m$}). This profile shows this region has a normalised intensity of { $I_{rs}/I_{a} = 0.56 \pm 0.06$} and a length of { $d = 850 \pm 100 \ \mu m$}. This yields a post-reverse shock mass density of { $40 \pm 11 \ mg/cc$}. However, it is possible that there is mixing between the neon and piston material in this region, which complicates the analysis due to different mass absorption cross section.

The mass density of the post-reverse shock region, $\rho_{rs}$, can also be estimated from the Rankin-Hugoniot equation for mass conservation $\rho_{1} u_{1} = \rho_{2} u_{2}$, where $\rho_{i}$ is the mass density and $u_{i}$ is the particle velocity in the shock frame. The initial shock moves with a velocity { $v_{s} = 80 \pm 10 \ km/s$} (measured in Section \ref{velocity}), into the unshocked gas, with a mass density { $ \rho_{a} = 0.49 \pm 0.01 \ mg/cc$}, and forms a post-shock region with a mass density $\rho_{ps}$ (estimated to be $12 \pm 1 \ mg/cc$ in Section \ref{post shock}). Particles in the unshocked gas have negligible initial velocities and so, in the shock frame, are moving at the shock velocity of $v_{s}$. Using the mass conservation equation, the velocity of particles in the post-shocked gas was calculated to be $\sim$$3 \ km/s$ in the shock frame. Subtracting this velocity from the shock velocity finds the post-shock particle velocity in the lab frame to be { $u_{ps} = 77 \pm 10 \ km/s$}.

\begin{equation}
	\centering
	u_{ps}= v_{s} - v_s  \frac{\rho_{a}}{\rho_{ps}} \sim 77 \pm 10 \ km/s
\end{equation}

The reverse shock propagates with a velocity { $v_{rs} = 35 \pm 12 \ km/s$} (measured in Section \ref{velocity}) into the post-shock region, forming a post-reverse shock region in the centre of the gas cell with a mass density $\rho_{rs}$. The post-reverse shock particles have stagnated at the gas cell centre with negligible velocity, and so in the reverse-shock frame move at the reverse shock velocity $v_{rs}$. Particles in the post-shock region, with lab frame velocity, $u_{ps}$, move towards the reverse shock at $u_{ps}+v_{rs}$. Using the mass conservation equation once again allows the mass density within the stagnated gas, $\rho_{rs}$, to be calculated to be { $38 \pm 10 \ mg/cc$}, agreeing with the mass density measured from XRBL images.

\begin{equation}
	\centering
	\rho_{rs} = \rho_{ps}  \frac{ u_{ps}+v_{rs} }{ v_{rs} } \sim 38 \pm 10 \ mg/cc
\end{equation}

The mass density within the post-reverse shock region was found to be { $38 \pm 10 \ mg/cc$}, corresponding to a compression of { $\times 76 \pm 12$} compared to the initial gas fill. However, the jump in density at the reverse shock front is only { $\times 3.2 \pm 0.6$}, below the compression limit of an ideal, monatomic gas of $\times 4$.

\subsection{The radiative precursor}
\label{precursor}
While XRBL and self-emission images can provide information about the post-shock region, they are not very sensitive to the radiative precursor ahead of the shock front. This region has a similar mass density to the unshocked gas, but is heated and ionized by radiation emitted from the shock front. Therefore, laser interferometry was used to measure the free electron density integrated along the probe beam path, $n_{e}L$ (units of $cm^{-2}$), which is related to the ionization of the unshocked gas.

\begin{figure*}[b]
	\includegraphics[width=\textwidth]{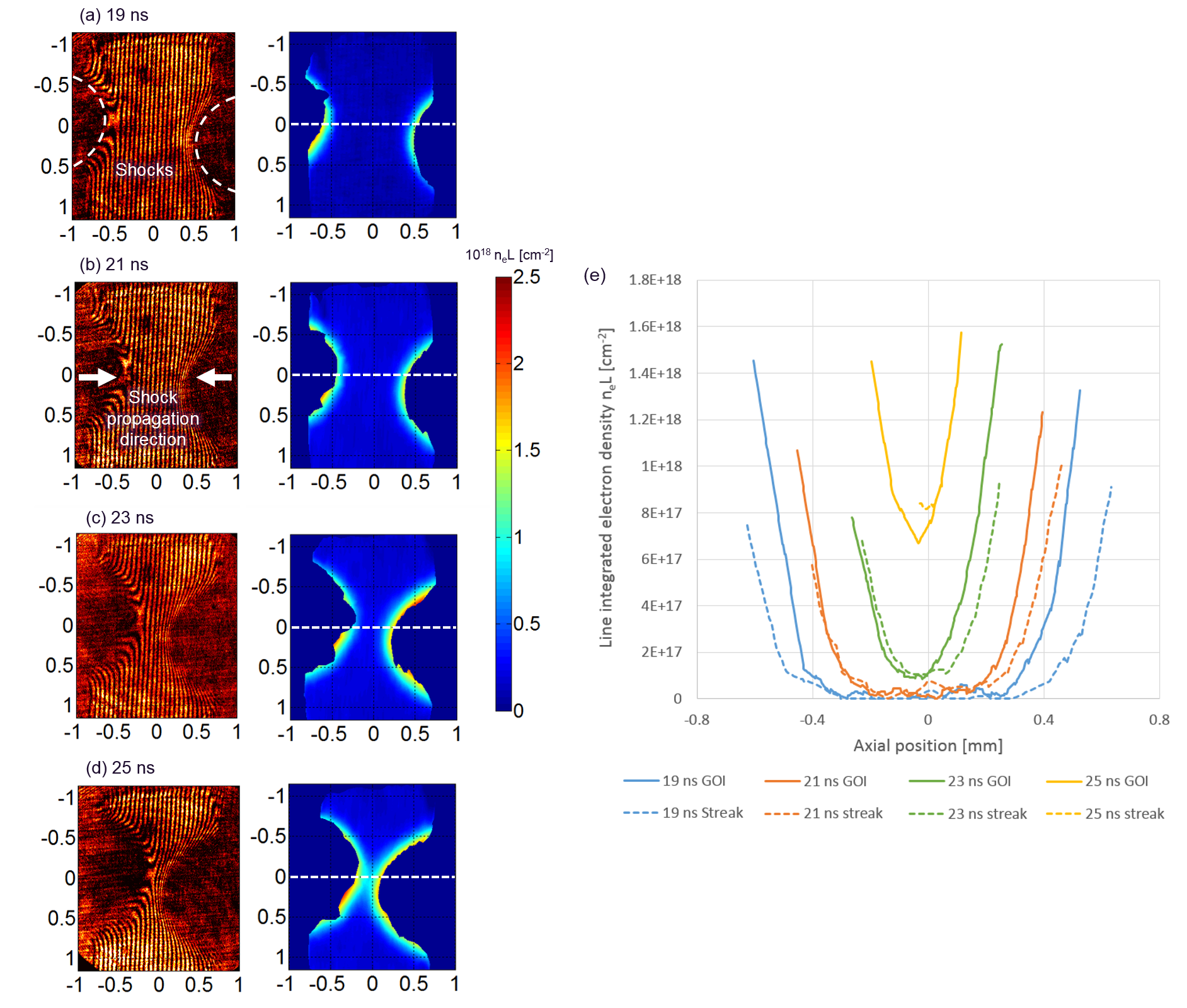}
	\caption{(a-d) Four GOI images from different times in the same shot, in neon. Left shows the raw interferograms and right shows the processed maps of line integrated electron density $n_{e}L$. (e) Axial lineouts plotted for all GOI images, compared to same time lineouts from streak interferometry (Fig \ref{fig:StreakInter}).}
	\label{fig:GOIs}
\end{figure*}
\FloatBarrier

The left hand side of Fig \ref{fig:GOIs}.a-d shows four Gated Optical Intensifier (GOI) images, all from the same shot in neon. The shocks appear as dark semi-circular shaped regions to the left and right of each image, indicated on Fig \ref{fig:GOIs}.a, similar to that seen in XRBL images in Fig \ref{fig:XRBL}. In contrast to XRBL images, the laser interferometry diagnostic does not need to be corrected for point projection scaling as the probe beam is collimated. This diagnostic directly measures changes in the refractive index integrated along the path of the probe beam. Shifts in the fringe position, from the original vertical lines, indicate a change in refractive index, and thus a free electron density and ionization in the unshocked plasma. This can be seen directly ahead of the shocked plasma in Fig \ref{fig:GOIs}.a-d, indicating the presence of a radiative precursor. The probe beam does not propagate through the shocked plasma or the region directly ahead of it due to large free electron densities, above the $532 \ nm$ probe beams critical density ($3.9 \times 10^{21} \ /cc$), and strong gradients in refractive index, which can refract the beam out of the collection optics.

Fringes in interferometry images, and corresponding reference images (with vertical fringes, not shown), were traced and processed as outlined by Swadling et al. \cite{Swadling2013}. The right hand side of Fig \ref{fig:GOIs}.a-d show the resulting maps of $n_{e}L$. These images show a smooth gradient in electron density outlining the shock front, indicating a gradual drop in ionization. The radiative precursor appears remarkably uniform, however the resolution of this diagnostic is limited to roughly half the inter-fringe spacing, $\sim$$32 \ \mu m$.

\begin{figure}
	\includegraphics[width=0.5\textwidth]{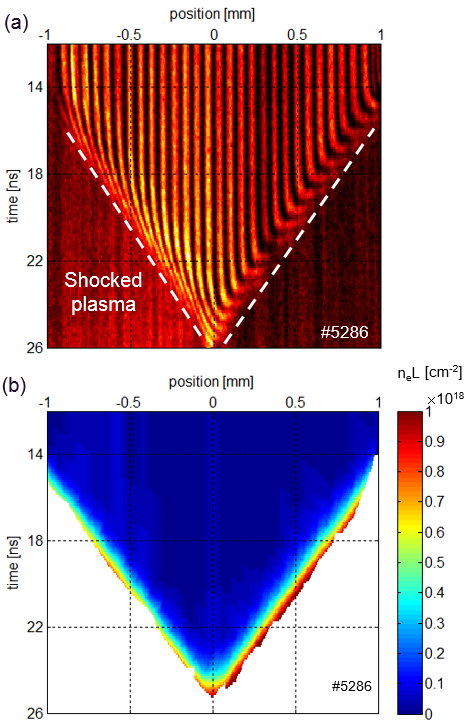}
	\caption{(a) Interferometry streak image along the axis for a shot in neon. Shifts in fringe position indicate changes in refractive index and thus a change proportional to the free electron density, $n_{e}$. The shocked plasma region was identified visually as the region opaque to the probe laser. (b) Processed map of line integrated electron density $n_{e}L$.}
	\label{fig:StreakInter}
\end{figure}

In addition to GOI cameras, a streak camera recorded interferometry along a 1-D line on the shock propagation axis. The original streak interferometry image (from the same shot as that in Fig \ref{fig:GOIs}) is shown in Fig \ref{fig:StreakInter}.a while the processed map of line integrated electron density, $n_{e}L$, is shown in Fig \ref{fig:StreakInter}.b. To compare the GOI and streak interferometry results from the same shot, 1-D lineouts were taken on axis of each GOI image and plotted in Fig \ref{fig:GOIs}.e as solid lines. 1-D lineouts taken from Fig \ref{fig:StreakInter}.b, at the times the GOI images, and plotted in Fig \ref{fig:GOIs}.e as dashed lines. The plots show very similar shapes, with similar distances between the shocks and gradients for the radiative precursors, indicating that the two diagnostics are consistent.

\begin{figure}
	\includegraphics[width=0.5\textwidth]{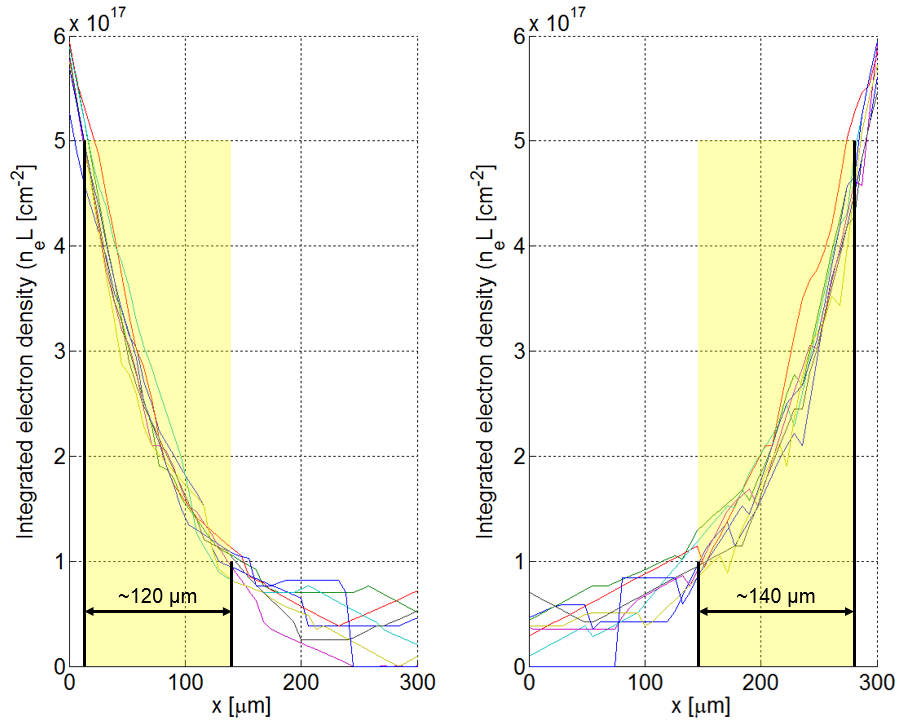}
	\caption{Lineouts of integrated electron density, from the map shown in Fig \ref{fig:StreakInter}.b, showing the radiative precursor ahead of the shock. Lineouts were taken $1 \ ns$ apart between $16 \ ns$ and $23 \ ns$ and shifted to intercept the y axis at a similar position. The width of the radiative precursor between $1 \times 10^{17} \ cm^{-2}$ and $5 \times 10^{17} \ cm^{-2}$ is reasonably constant at $130 \pm 10 \ \mu m$.}
	\label{fig:ConstantPrec}
\end{figure}

Fig \ref{fig:ConstantPrec} shows lineouts of $n_{e}L$ at constant times from Fig \ref{fig:StreakInter}.b. Lineouts were taken every $1 \ ns$ between $16 \ ns$ and $23 \ ns$ and shifted to intercept the y axis at a similar position. These lineouts follow a similar shape suggesting that the radiative precursor remains constant over $7 \ ns$ (the time the shocks are visible within the target window before the radiative precursors interact). The spatial width of the radiative precursor  can be measured between $1 \times 10^{17} \ cm^{-2}$ and $5 \times 10^{17} \ cm^{-2}$, selected to be above the noise floor and below the point where the $532 \ nm$ probe beam does no propagate through the plasma. This yielded a width of $130 \pm 10 \ \mu m$ for both the left and the right shocks

A constant radiative precursor width implies that the radiative flux from the shock, and thus the post shock temperature, is approximately constant over the observed time. A rough estimate for the post shock temperature can then obtained by equating the radiative loss flux, $F_{R}$, to the kinetic energy flux of incoming particles, $F_{KE}$. From arguments outlined by Drake et al. \cite{Drake2011}, the radiative flux for a post-shock region in thermal equilibrium can be approximated as twice the blackbody emission, where $\sigma_{SB}$ is Stefan Boltzmann constant and $T$ is the temperature. The kinetic energy flux of incoming particles can be approximated as half the product of the unshocked mass density $\rho_{a}$ and the shock velocity (measured to be { $80 \pm 10\ km/s$} in Section \ref{velocity}) cubed, $v_{s}^{3}$.

\begin{equation}
	\centering
	\begin{split}
		F_{R} \sim F_{KE}
		\\
		2 \sigma_{SB} T^{4} \sim \frac{1}{2} \rho_{a} v_{s}^{3}
	\end{split}
\end{equation}

This model approximates the post-shock temperature to be { $15.7 \pm 1.7 \ eV$} with a similar result is found for shots in other gases at similar mass densities (Section \ref{different gases}). However, this simple model does not account for losses to ionizing, $F_{I}$, and heating, $F_{TH}$, the incoming flux of particles and thus may overestimates the temperature. The model may be slightly improved by balancing the incoming kinetic energy flux against the losses to radiation, ionization and heating.

\begin{equation}
\centering
F_{KE} = F_{R} + F_{I} + F_{TH}
\label{eq:energy}
\end{equation}

The loss to ionizing the incoming flux of particles, $F_{I}$, can be approximated as the incoming flux of particles, $n$ ($n = \rho_{a} v_{s} / A m_{p}$ where $A$ is the atomic mass and $m_{P}$ is the proton mass), multiplied by the ionization potential for the average final ionization state, $\bar{Z}$. The average final ionization state was determined using the commercially available software PrismSPECT \cite{MacFarlane2006}. A $100 \ \mu m$ planar slab of neon plasma was simulated at average densities found in the post-shock region, $12 mg/cc$, for a variety of possible temperatures. The ionization potential per ion was then determined with data from \cite{NIST_ASD}. The loss to heating the incoming particle flux was approximated as the incoming flux of particles multiplied by the internal energy of a monatomic ideal gas $F_{TH} = \frac{3}{2} n (1+\bar{Z}) k_{B} T$, where the $(1+ \bar{Z} )$ accounts for heating of electrons assuming the electrons and ions are thermalised. Solving Eq \ref{eq:energy} yields a slightly lower post-shock temperature of  { $14.4 \pm 1.6 \ eV$}. This model suggests that in neon { $12 ^{+2}_{-1} \%$} of the incoming kinetic energy is used to heat the incoming atoms, { $17 ^{+4}_{-2} \%$} is lost to ionizing them and the remaining { $71 ^{+7}_{-3} \%$} is radiated upstream. However, simulations (detailed in Section \ref{sims}) suggest that higher temperatures are reached in the post-shock material and so a dedicated diagnostic for measuring temperature would be required for future experiments.

\subsection{The radiative precursor in different gases}
\label{different gases}
\begin{figure*}[t]
	\includegraphics[width=\textwidth]{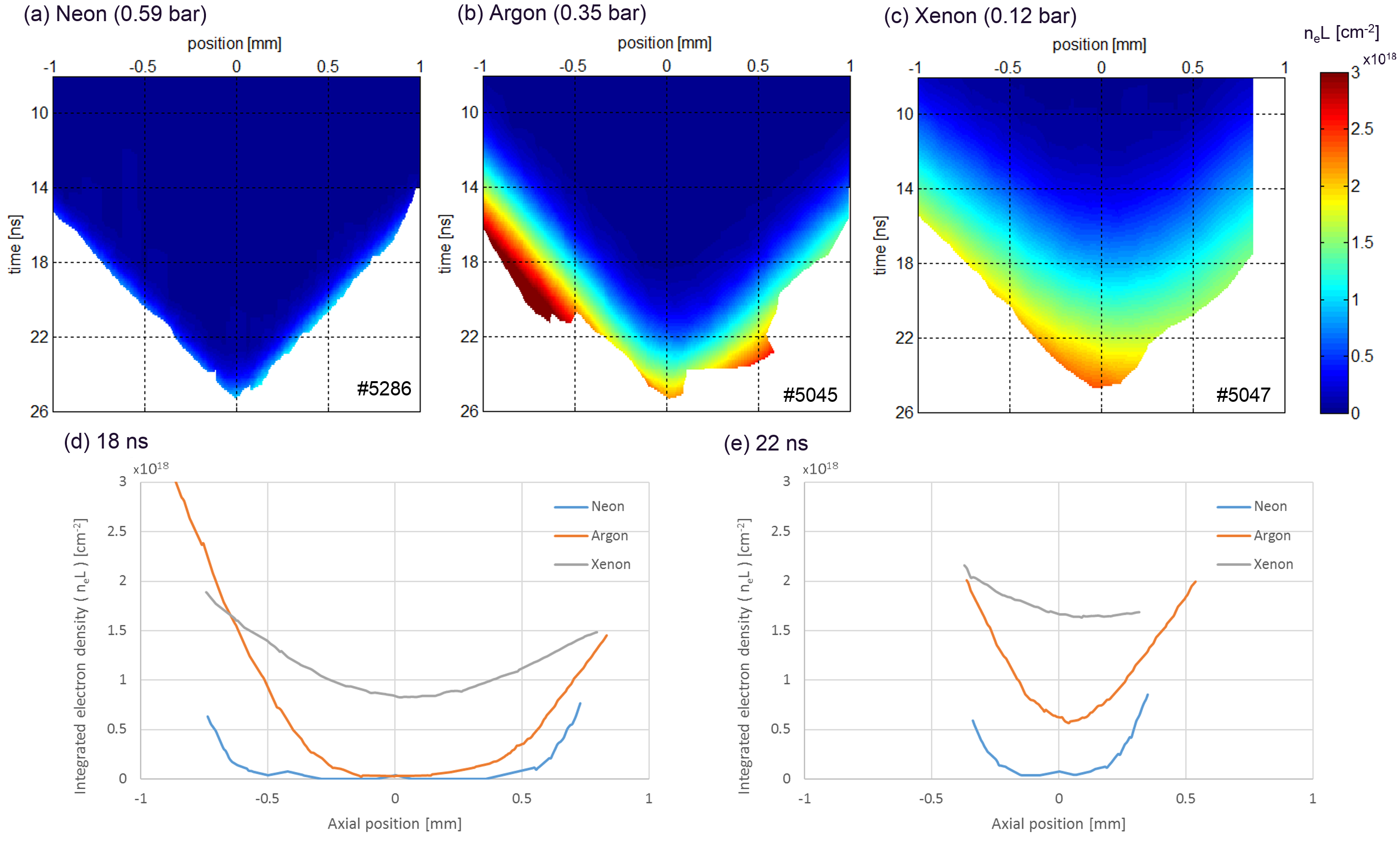}
	\caption{(a,b,c) Interferometry streak images along the axis for shots in different gases but similar initial mass densities, $0.56 \pm 0.08 \ mg/cc$. (a) neon at $0.48 \ mg/cc$ ($0.59 \ bar$), (b) argon at $0.56 \ mg/cc$ ($0.35 \ bar$), (c) xenon at $0.64 \ mg/cc$ ($0.12 \ bar$). The radiative precursor in heavier elements extends significantly further downstream. (e,d) 1-D lineouts of integrated electron density at constant times. (d) $18 \ ns$, (e) $22 \ ns$.}
	\label{fig:PrecGases}
\end{figure*}

These experiments also investigated the formation of radiative precursors in several different gases at similar initial mass densities. This ensured that the shocks evolved with similar hydrodynamics and thus similar shock velocities. Fig \ref{fig:PrecGases}.a-c show maps $n_{e}L$, from streak images in three different shots with different gases (neon, argon and xenon) with similar initial mass densities ($0.58 \pm 0.08 \ mg/cc$). In addition, Fig \ref{fig:PrecGases}.d-e show 1-D lineouts from these maps (at a constant time) plotted for all three gases, $18 \ ns$ and $22 \ ns$.

Heavier elements have a visibly larger precursor. The width of the radiative precursor was measured between constant values of $n_{e}L$ as in Section \ref{precursor} and the results are summarised in Table \ref{table:PrecGases}. In addition the interferometery diagnostic is able to probe to higher values of $n_{e}L$ in heavier elements. This is due to smoother gradients in free electron density, resulting in light not being refracted out of the collection optics. The radiative precursor in higher atomic number elements extends significantly further downstream, almost four times further in xenon compared to neon. This is likely due to lower ionization energy of outer shell electrons in heavier elements. In addition, as was found with the neon case, the width of the radiative precursor, in the argon and xenon cases, remains approximately constant over the observed period. A similar argument can therefore be made, as that found in Section \ref{precursor}, to estimate the post shock temperature to be $ \sim$$15 \ eV$. Further comparisons between radiative shocks in different gases and pressures will be the subject of future work.

\begin{table*}[t]
	\centering
	\begin{tabular}{|c|c|c|c|}
		\multicolumn{1}{c}{ \textbf{Gas} }
		&  \multicolumn{1}{c}{ \textbf{Density} }
		&  \multicolumn{1}{c}{ \textbf{Pressure} }
		& \multicolumn{1}{c}{ \textbf{Precursor Width} } \\
		\hline
		
		Neon (Z=10) & $0.48 \ mg/cc$ & $0.59 \ bar$ & $130 \pm 20 \ \mu m$ \\
		\hline
		Argon (Z=18) & $0.56 \ mg/cc$ & $0.35 \ bar$ & $220 \pm 30 \ \mu m$ \\
		\hline
		Xenon (Z=54) & $0.64 \ mg/cc$ & $0.12 \ bar$ & $500 \pm 50 \ \mu m$ \\
		\hline
	\end{tabular}
	\caption{Summary of precursor spatial width (measured from Fig \ref{fig:PrecGases}) between $n_{e}L$ values of $1 \times 10^{17} \ cm^{-2}$ and $5 \times 10^{17} cm^{-2}$ (selected to be above the noise floor and bellow the point where the laser does no propagate through the plasma for all images) for different gas fills at a similar initial mass density ($0.58 \pm 0.08 \ mg/cc$).}
	\label{table:PrecGases}
\end{table*}

\section{Simulation results}
\label{sims}
In this section, results from 1 and 2-dimensional radiation hydrodynamic simulations, conducted before and after the experiments, are presented. Simulation results are then compared to shock quantities measured and estimated from experiments in the previous section.

\subsection{NYM/PETRA 2-D simulations}
\label{2D sims}
Prior to the experiment, 2-dimensional radiative-hydrodynamic simulations were performed to estimate conditions and dynamics of the plasma and help determine XRBL target materials. These simulations assumed cylindrical symmetry and identical colliding shocks.

The first few nanoseconds (e.g. $5 \ ns$), including the laser-piston interaction, were simulated with the Lagrangian code NYM \cite{Roberts1980}. This used multi-group implicit Monte-Carlo X-ray transport and full laser-interaction physics. To avoid mesh tangling, the evolution and shock dynamics were then simulated with the Eulerian code PETRA \cite{Youngs1984}, which included multi-group X-ray diffusion. Equations of state and opacities for the piston materials and gas fill were taken from SESAME tables. Simulations assumed optimal Orion firing conditions with a spot size of $600 \ \mu m$ and $2 \ kJ$ drive beam energy ($30\%$ higher than that typically achieved in experiments,  $ \sim$$1.52 \ kJ$), nevertheless achieving qualitatively similar results.

\begin{figure}
	\includegraphics[width=0.5\textwidth]{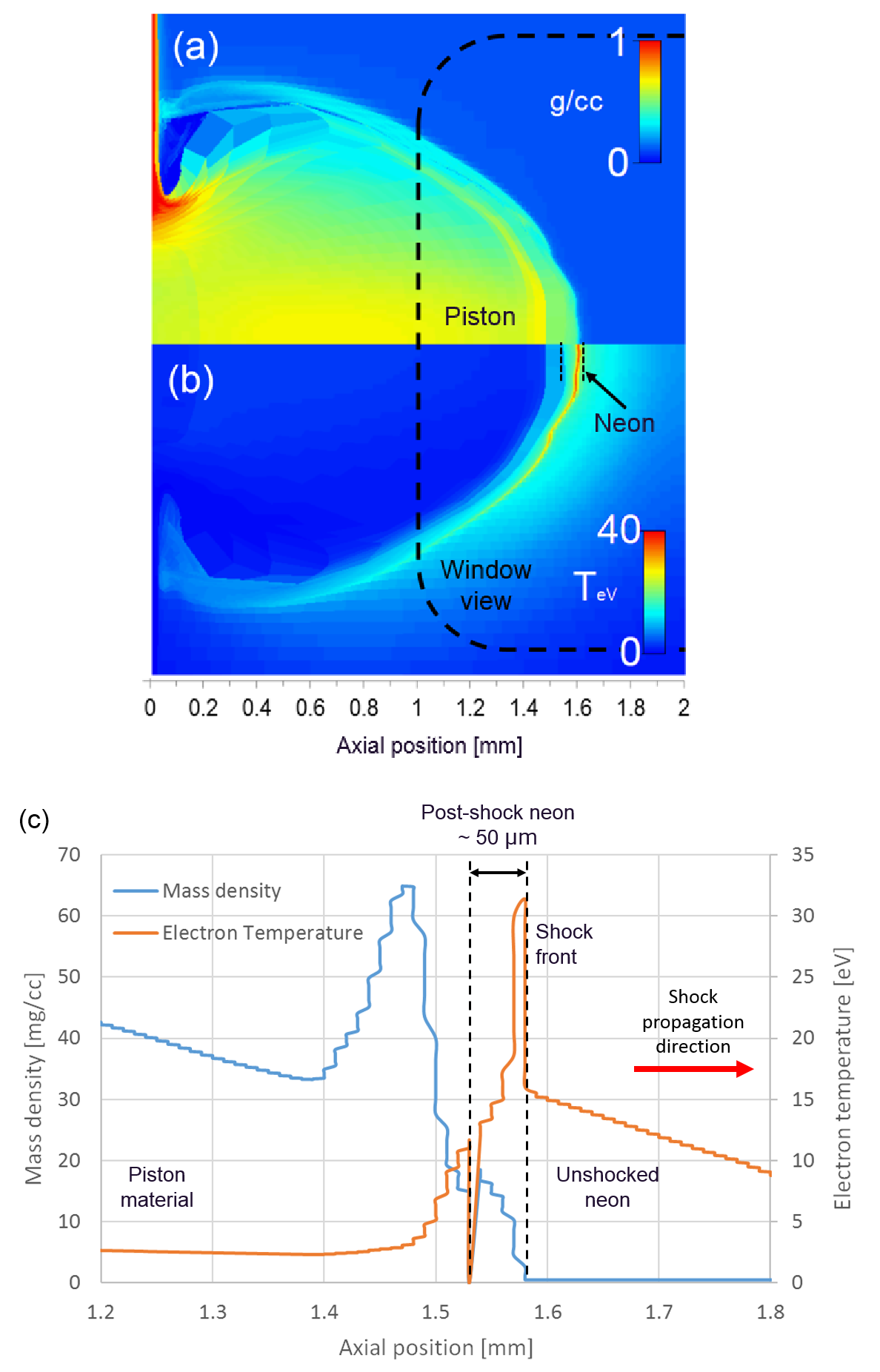}
	\caption{Numerical simulations of shock formation in neon at $0.5 \ mg/cc$ for maximum Orion drive energy ($2 \ kJ$ and $600 \ \mu m$ laser spot size). 2-D plots, at $16 \ ns$, of (a) mass density (in $g/cc$, log scale), (b) electron temperature (in $eV$, linear scale). The dashed line represents the view through the target diagnostic windows. 1-D lineouts on axis of (c) plots of mass density and electron temperature through the axis from $1 \ mm$ to $2 \ mm$.}
	\label{fig:2Dsims}
\end{figure}

Fig \ref{fig:2Dsims}.a-b shows 2-D maps of mass density (log scale) and electron temperature (linear scale) of the numerical simulations at $16 \ ns$. Lineouts, along the axis of shock propagation, for mass density and electron temperature are shown in Fig \ref{fig:2Dsims}.c. At $16 \ ns$, simulations show a very thin $ \sim$$50 \  \mu m$ region of shocked neon, $1.54 \ mm$ from the piston, {with an average mass density of $15.1 \ mg/cc$} (corresponding to a maximum post-shock compression of {$\times 30.2$}) and a post shock temperature of {$12.0-16.0 \ eV$ (excluding the spike in temperature at the shock front)}.

\subsection{HELIOS 1-D simulations}
\label{1D sims}
\begin{figure}
	\includegraphics[width=0.5\textwidth]{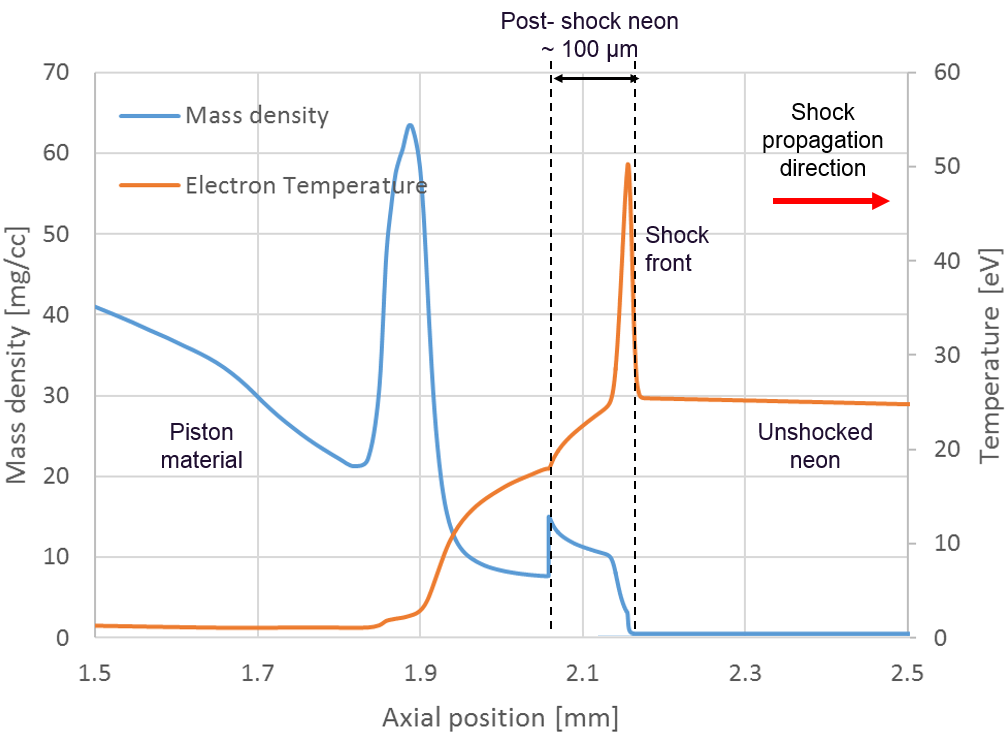}
	\caption{Axial plots of mass density (in $mg/cc$) and electron temperature (in $eV$). The dashed line represents the region of post-shock neon.}
	\label{fig:HeliosPlot} 
\end{figure}

After the experiments, 1-dimensional radiative-hydrodynamic, multi-group simulations were performed with the commercial available software HELIOS \cite{MacFarlane2006}. Fig \ref{fig:HeliosPlot} shows 1-D plots of both mass density and electron temperature at $16 \ ns$. These simulated a single shock being driven into neon at $0.5 \ mg/cc$ by a $6 \times 10^{14} \ W/cm^{2}$ laser (similar to that achieved in experiments). Both layers of the piston were simulated as being composed of only CH but with mass densities similar to those used in the experiments.

At $16 \ ns$, these simulations also show a thin $ \sim$$100 \  \mu m$ region of shocked neon, $2.06 \ mm$ from the piston, {with an average mass density of $10.5 \ mg/cc$} (corresponding to a maximum post-shock compression of {$\times 20.9$}) and a post-shock electron temperature of {$17.9-25.4 \ eV$ (excluding the spike in temperature at the shock front)}. Despite lower drive energies, the 1-D HELIOS simulations achieve a higher post-shock electron temperature than the 2-D NYM/PETRA simulations.

\subsection{Comparison between simulations and experiments}
\label{sims compare}
The results of both 1-D and 2-D simulations were compared against experimentally measured and estimated quantities, and summarised in Table \ref{table:SimCompar}. Both 1-D and 2-D simulations achieve similar velocities and post-shock conditions to those estimated for experiments in neon at $0.49 \pm 0.01 \ mg/cc$.

%
%

\begin{table*}[t]
	\centering
	\begin{tabular}{c|c|c|c|}
		\multicolumn{1}{r}{}
		&  \multicolumn{1}{c}{ \textbf{Experimental} }
		& \multicolumn{1}{c}{ \textbf{NYM-PETRA (2-D)} }
		& \multicolumn{1}{c}{ \textbf{HELIOS (1-D)} } \\
		\cline{2-4}
		
		\textbf{Shock velocity} & { $80 \pm 10 \ km/s$} & $92 \ km/s$ & $116 \ km/s$ \\
		\cline{2-4}
		{\textbf{Average post-shock compression}} & { $\times 25 \pm 2$} & {$\times 30.2$} & {$\times 20.9$} \\
		\cline{2-4}
		\textbf{Post-shock temperature} & { $14.4 \pm 1.6 \ eV$} & {$12.0 - 16.0 \ eV$} & {$17.9 - 25.4 \ eV$} \\
		\cline{2-4}
	\end{tabular}
	
	\caption{Summary of shock parameters estimated from experimental data, compared to both 1-D and 2-D radiation hydrodynamic simulations.}
	\label{table:SimCompar}
\end{table*}

The 1-D HELIOS simulation overestimates the velocity and post shock temperature, {but predict similar average post shock compressions to that seen in the experiment}. The 2-D NYM-PETRA simulations were run with $30\%$ higher drive energy and so result in slightly higher velocities and significantly higher post-shock compressions. However, despite the higher drive energy, {these simulations predict a post-shock electron temperature in-line with that estimated with experimental data and simple models.} The temperature estimates for these experiments were determined indirectly and relied on several assumptions, and so a dedicated diagnostic for measuring post-shock temperature would be required for future experiments.

\section{Discussion and Summary}
\label{summary}
The experiments detailed in this paper introduced a new experimental configuration to study the formation of radiative shocks expanding freely into large gas volumes and, for the first time, study the interaction of two counter-propagating radiative shocks. The Orion high-power laser facility, at AWE Aldermaston UK, was used to produce counter-propagating radiative shocks in neon, argon and xenon with initial pressures between $0.1$ to $1.0 \ bar$. A combination of optical self-emission, laser interferometry and X-ray backlighting (XRBL) allowed the shock structure and radiative precursor to be studied simultaneously.

With an initial gas fill of neon at $0.49 \pm 0.01 \ mg/cc$, shocks propagated with velocities of { $80 \pm 10 \ km/s$} and experienced strong radiative cooling resulting in post-shock compression of { $\times 25 \pm 2$}. After the collision of the counter-propagating shocks, reverse shocks were formed with a compression of { $\times 76 \pm 12$}. Simple models were used to estimate the post shock temperature to be { $14.4 \ \pm 1.6 \ eV$}. Finally, three gases (Ne, Ar and Xe) with similar initial mass densities ($0.58 \pm 0.08 \ mg/cc$) were compared. Although they exhibited similar shock dynamics, the radiative precursor was found to extend significantly further downstream in hight atomic number gases ($\sim$$4$ times further in xenon compared to neon). The experimental data is in reasonable agreement with 1-D and 2-D radiative-hydrodynamic simulations and provides a new benchmark for additional codes to be tested against.

Further work is required to understand the dynamics of the colliding shocks and determine the impact of radiative effects on the resulting complex structure. Density perturbations in the shock front, prior to collision (labelled $\lambda$ in Fig \ref{fig:XRBL}.a), are visible on XRBL images and measured to be $90 \pm 20 \ \mu m$ in width. The nature of these will be the subject of future work but may suggest the formation of hydrodynamic or radiative instabilities.

\FloatBarrier

\section*{Acknowledgements}
This work was supported by STFC and AWE through Orion's Academic Access Programme, by The Royal Society and EPSRC through a DTA studentship and Labex PLAS@PAR (ANR-11-IDEX-0004-02). The authors would like to acknowledge Robert Charles, Jim Firth, Paul Treadwell, Rob Johnson, David Hillier, Nick Hopps and the entire Orion team at AWE Aldermaston for their help and support during the experiments.

\bibliographystyle{ieeetr}
\bibliography{MendeleyOutput,otherBib}

\end{document}